\documentclass[10pt,aps,twocolumn, pra, footinbib,superscriptaddress]{revtex4-1}
\usepackage{graphicx}
\usepackage{psfrag}
\usepackage{amsmath,amssymb}
\usepackage{colordvi}
\usepackage{color}
\usepackage{hyperref}
\usepackage{float}
\hypersetup{colorlinks=true, linktoc=all, linkcolor=blue, urlcolor=blue,citecolor=red}

\begin{document}

\title{Three-Body Problem of Bosons nearby a $d$-wave Resonance}
\author{Juan Yao}
\affiliation{Institute for Advanced Study, Tsinghua University, Beijing, 100084, China}

\author{Pengfei Zhang}
\affiliation{Institute for Advanced Study, Tsinghua University, Beijing, 100084, China}

\author{Ran Qi}
\affiliation{Department of Physics, Renmin University of China, Beijing, 100872, P. R. China}

\author{Hui Zhai}
\affiliation{Institute for Advanced Study, Tsinghua University, Beijing, 100084, China}
\affiliation{Collaborative Innovation Center of Quantum Matter, Beijing, 100084, China}

\date{\today}

\begin{abstract}
Motivated by recent experimental progresses, we investigate few-body properties of interacting spinless bosons nearby a $d$-wave resonance. Using the Skorniakov-Ter-Martirosion (STM) equations, we calculate the scattering length between an atom and a $d$-wave dimer, and we find that the atom-dimer scattering length is positive and is much smaller the result from the mean-field approximation. We also reveal unique properties of the three-body recombination rate for a degenerate Bose condensate nearby the $d$-wave resonance. We find that the total recombination rate is nearly a constant at the quasi-bound side, in contrast to the behavior of a thermal gas nearby high-partial wave resonance. We also find that the recombination rate monotonically increases across the unitary point toward the bound side, which is due to the largely enhanced coupling between the atom and the $d$-wave dimer with deeper binding energy. This monotonic behavior is also qualitatively different from that of a degenerate gas nearby an $s$-wave resonance counterpart. 

\end{abstract}

\maketitle

\section{Introduction} 

Recently broad $d$-wave shape resonances have been observed in Bose gases of ${}^{85}$Rb-${}^{87}$Rb mixture \cite{YCui} and of ${}^{41}$K atoms \cite{XCYao}. Since the resonances are broad, it is much easy for experiments to locate the magnetic field stably at the resonance regime. In addition, the lifetime of the Bose gas nearby the $d$-wave resonance is found to be much longer than the many-body time scale \cite{XCYao}. These properties enable studying novel many-body physics of a $d$-wave resonant interacting Bose gas. In fact, certain many-body effects have been observed in a recent experiment when a degenerate Bose gas is ramped through the $d$-wave resonance \cite{XCYao}.

So far the few-body properties of this system have been barely studied \cite{JiaWang, PF SZ ZH, JYd}. However, these few-body properties are quite important for understanding this system. For instance, the scattering length between an atom and a d-wave dimer is crucial for the stability of an atom-dimer mixture produced through ramping through the resonance, as well as for determining the system size and its collective mode. The three-body recombination rate is also crucial for understanding the lifetime of this system. 

In this work, we study the three-boson problem nearby a $d$-wave resonance and with a background $s$-wave interaction, using the Skorniakov-Ter-Martirosion (STM) equations \cite{Braaten}. We extract the scattering length between an atom and a d-wave dimer and the three-body recombination rate from the calculation. Our main findings are:

i) We find that the scattering length between an atom and a $d$-wave dimer is positive and is generally smaller than $8a_s/3$ with $a_s$  being the the atom-atom $s$-wave scattering length. Here $a_{ad}^{\rm MF}=8a_s/3$ is the atom-dimer scattering length estimated by the mean-filed approximation. 

ii) The three-body recombination rate is nearly a constant at the quasi-bound state and monotonically increases across the unitary regime toward the bound side. We point out that this behavior is qualitatively different from both a thermal gas nearby high-partial wave resonance and a degenerate gas nearby an $s$-wave resonance. 

\section{The Model} 

For a $d$-wave resonantly interacting spinless Bose system with background $s$-wave interaction, the Lagrangian of the effective field theory can be casted as  
\begin{equation}\begin{aligned}
\mathcal{L}&=\sum_{\bf k}\psi^\dagger_{\bf k}\left(i\partial_t-\frac{k^2}{2M}\right)\psi_{\bf k}
+\sum_{\bf q}\frac{\bar{\nu}}{\bar{g}_d^2}d^\dagger_{\bf q} d_{\bf q}
+\sum_{\bf q}\frac{1}{\bar{g}_s}\phi^\dagger_{\bf q}\phi_{\bf q}\\
&-\sum_{\bf k, q}\left[\frac{\sqrt{2\pi}k^2Y_{20}(\hat{k})e^{-k^2/\bar{\Lambda}_d^2}}{\sqrt{V}} d^\dagger_{2{\bf q}} \psi_{{\bf q}+{\bf k}/2}\psi_{{\bf q}-{\bf k}/2}+{\rm H.c.}\right] \\
&-\sum_{\bf k, q}\left[\frac{e^{-k^2/\bar{\Lambda}_s^2}}{\sqrt{V}} \phi^\dagger_{2 {\bf q}}\psi_{{\bf q}+{\bf k}/2}\psi_{{\bf q}-{\bf k}/2}+{\rm H.c.}\right], \label{model}
\end{aligned}\end{equation}
where $\psi(^{\dagger})$ is the atomic field, and $d(^{\dagger})$ is the $d$-wave dimer field. Here $\phi$ is the $s$-wave dimer field  introduced through the Hubbard-Stratonovich transformation to incorporate the $s$-wave interaction between atoms, and the last line describes the conversion between atoms and $s$-wave dimers, where the form factors $e^{-k^2/
\bar{\Lambda}_{s(d)}^2}$ are introduced to regularize the ultraviolet behavior. The atoms and $d$-wave dimers are coupled through the second line of the above equation. A separable potential is used for both the $d$-wave and $s$-wave interactions. In principle, the $d$-wave dimers have five components corresponding to $m=0,\pm 1,\pm 2$, but they are split due to the dipole-dipole interaction. Here, without loss of generality, we will only consider the $m=0$ channel with coupling characterized by the spherical harmonic function $Y_{20}(\hat{k})$. 

\begin{figure}[t]
\begin{centering}
\includegraphics[width=0.45\textwidth]{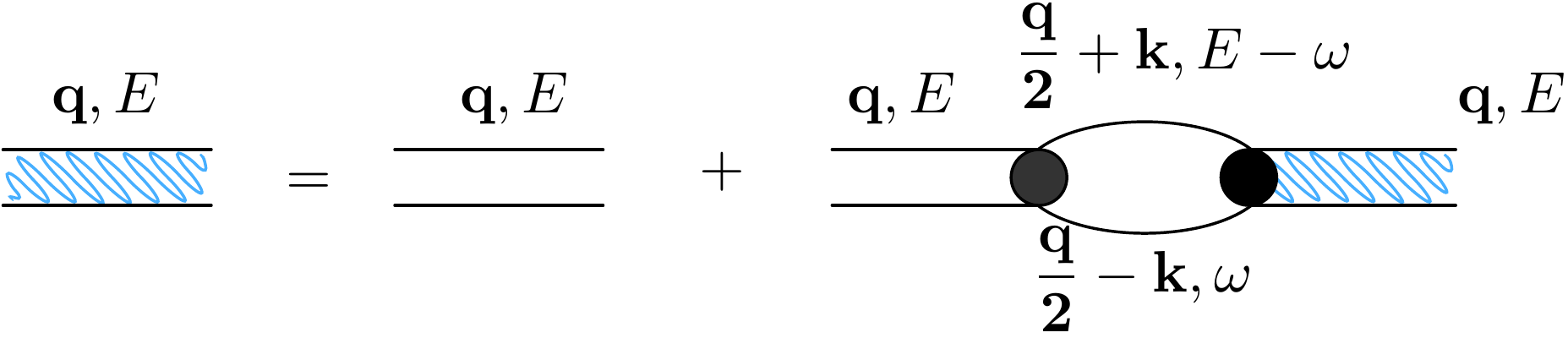}
\caption{Feynman rules for the $s$-wave dimer full propagator $iD_s({\bf q}, E)$. The single line stands for the bare atomic Green's function. The double lines are the bare $s$-wave dimer Green's function. The black dots are the $s$-wave interacting vertex.}
\label{Fig1}
\end{centering}
\end{figure}

The parameters in the model Eq. \ref{model} needs to be related to physical parameters. First of all, the bare parameters relevant to the $s$-wave interaction are $\bar{g}_s$ and $\bar{\Lambda}_s$. These bare parameters are determined to reproduce the correct physical parameters of scattering length $a_s$ and bound state energy $E_b^s={-5.7}/(2Ma_\text{s}^2)$ which is calculated for $^{41}$K by the quantum-defect theory. Here the reason we use the binding energy from the quantum defect theory is because the $s$-wave scattering length is generically small nearby the $d$-wave resonance, as in the case of these two experiments \cite{YCui,XCYao}, and therefore the result from a zero-range potential $E_b=-1/(Ma^2_\text{s})$ is not quite accurate. 

The $s$-wave scattering matrix can be derived as 
\begin{equation}
T_s\left(k\hat{k},k\hat{k}',E=\frac{k^2}{M}\right)=-\frac{4e^{-2k^2/\bar{\Lambda}_s^2}}{V}D_s({\bf q}=0,E=\frac{k^2}{M}),
\label{EqTs}
\end{equation}
where $D_s({\bf q},E)$ is the $s$-wave dimer full propagator which can be diagrammatically described by Fig. \ref{Fig1},
\begin{equation}\begin{aligned}
&D_s^{-1}({\bf q},E)\equiv D_s^{-1}(\Omega=E-\frac{q^2}{4M})\\
&=\frac{M}{2\pi a_s}-\frac{M\sqrt{-M\Omega}}{2\pi}\left(1-\text{Erf}\left[\frac{\sqrt{-2M\Omega}}{\Lambda_s}\right]\right)e^{-\frac{2M\Omega}{\bar{\Lambda}_s^2}},
\label{EqDs}
\end{aligned}\end{equation}
where $\text{Erf}(x)$ is the Error function defined through $\text{Erf}(x)=\frac{2}{\sqrt{\pi}}\int_0^x dt e^{-t^2}$. Here the bare parameter $\bar{g}_s$ has been regularized through 
\begin{equation}
\frac{1}{\bar{g}_s}=\frac{M}{2\pi a_s}-\frac{M}{\pi^2}\int d k e^{-2k^2/\bar{\Lambda}_s^2}.
\label{EqsNorgs}
\end{equation}
The above renormalization relation is established through the connection between the $s$-wave scattering matrix Eq. \eqref{EqTs} and the s-wave phase shift
\begin{equation}
T^{-1}_s\left(k\hat{k},k\hat{k}',E=\frac{k^2}{M}\right)=-\frac{VM}{8\pi}\left({\cot\delta_s-ik}\right).
\end{equation}
Taking the low energy expansion of the $s$-wave phase shift up to zeroth order of $k$ with $\cot\delta_s=-1/a_s$, renormalization relation between $\bar{g}_s$ and $a_s$ is then established as in Eq. \eqref{EqsNorgs}.

On the other hand, the bound state energy can be calculated by requiring $D_s^{-1}({\bf q}={\bf 0},E=E^s_b)=0$. The binding energy obtained in this way is required to be equal to $E_b^s={-5.7}/(2Ma_\text{s}^2)$ obtained from the quantum defect theory, which leads to the following renormalization between $a_s$ and $\bar{\Lambda}_s$ as 
\begin{equation}
\bar{\Lambda}_s=\frac{4.36}{a_s}.
\end{equation}

Secondly, the bare parameters relevant to the $d$-wave interaction are $\bar{g}_d$, $\bar{\Lambda}_d$ and $\bar{\nu}$. Similar to diagrams in Fig. \ref{Fig1}, the $d$-wave full propagator is calculated as 
\begin{equation}\begin{aligned}
&D^{-1}_d({\bf q},E)\equiv D^{-1}_d(\Omega=E-\frac{q^2}{4M}) \\
&=\frac{\bar{\nu}}{\bar{g}_d^2}+
\frac{\sqrt{2\pi}M\bar{\Lambda}_d[16{(M\Omega )}^2+4{M\Omega }\bar{\Lambda}_d^2+3\bar{\Lambda}_d^4]}{128\pi^2}\\
&-\frac{32M(-M\Omega)^{5/2}e^{-\frac{2M\Omega}{\bar{\Lambda}_d^2}}\text{Erfc}\left(\frac{\sqrt{-2M\Omega}}{\bar{\Lambda}_d}\right)}{128\pi^2},
\label{EqDd}
\end{aligned}\end{equation}
where the complementary error function $\text{Erfc}(x)$ is defined as $\text{Erfc}(x)=1-\text{Erf}(x)$. Then the associated $T$-matrix is given as
\begin{equation}\begin{aligned}
T_d(k\hat{k},k\hat{k}',E=\frac{k^2}{M})=-\frac{8\pi k^4 Y_{20}(\hat{k})Y_{20}(\hat{k}') e^{-2\frac{k^2}{\bar{\Lambda}_d^2}}}{V}&\\
\times D_d({\bf q}=0,E=\frac{k^2}{M})&,
\end{aligned}\end{equation}
from which one can obtain the two-body scattering amplitude in the $d$-wave channel as
\begin{equation}
T_d(k\hat{k},k\hat{k}',E=\frac{k^2}{M})=-\frac{32\pi^2}{MV}Y_{20}(\hat{k})Y_{20}(\hat{k}')f_d(k).
\end{equation}
The $d$-wave scattering amplitude $f_d(k)$ is related to the $d$-wave phase shift through
\begin{equation}
f_d(k)=\frac{k^4}{k^5\cot\delta_d(k)-ik^5}.
\end{equation}
In the low energy limit, the $d$-wave phase can be expanded in order of $k$ as $k^5\cot\delta_d=-1/D-k^2/v-k^4/R$. With these one can establish the following three renormalization relations between the bare $d$-wave interacting parameters $\{\bar{\nu}, \bar{g}_d, \bar{\Lambda}_d\}$ and the physical parameters $\{ D, v, R\}$,
\begin{equation}\begin{aligned}
\frac{1}{D}&=\frac{4\pi}{M}\frac{\bar{\nu}}{\bar{g}_d^2}+\frac{3}{16\sqrt{2\pi}}\bar{\Lambda}_d^5 \\
\frac{1}{v}&=\frac{8\pi}{M\bar{\Lambda}_d^2}\frac{\bar{\nu}}{\bar{g}_d^2}+\frac{5}{8\sqrt{2\pi}}\bar{\Lambda}_d^3 \\
\frac{1}{R}&=\frac{8\pi}{M\bar{\Lambda}_d^4}\frac{\bar{\nu}}{\bar{g}_d^2}+\frac{15}{8\sqrt{2\pi}}\bar{\Lambda}_d.
\label{EqNormald}
\end{aligned}\end{equation}
In the following calculation, for each set of $\{ D, v, R\}$, we numerically solve the $\{\bar{\nu}, \bar{g}_d, \bar{\Lambda}_d\}$.

\begin{figure}[t]
\begin{centering}
\includegraphics[width=0.45\textwidth]{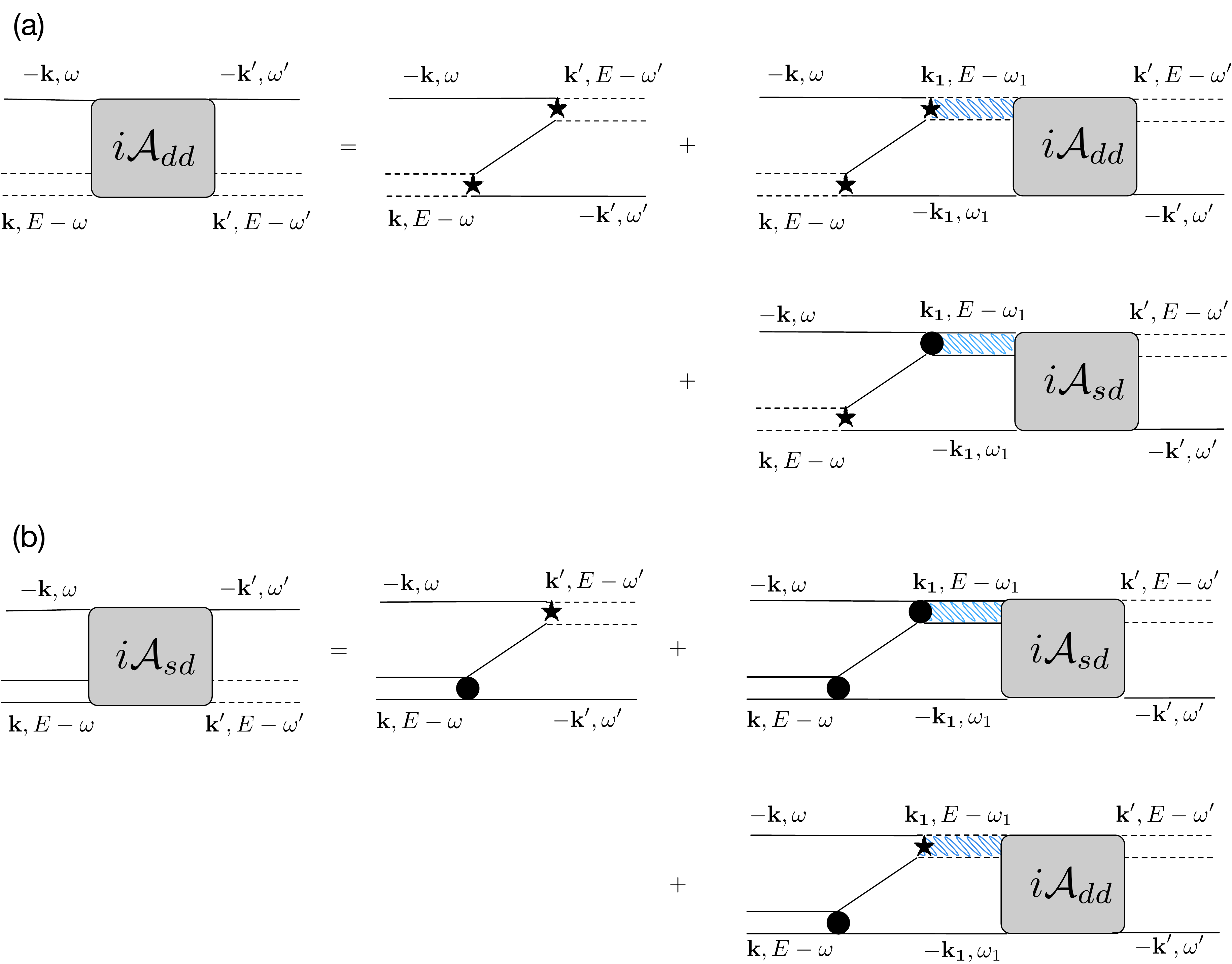}
\caption{Feynman rules for the scattering processes between an atom and a $d$-wave dimer. The double (dashed) lines are the bare $s$ ($d$)-wave dimer Green's functions. The shaded double lines are the corresponding full Green's functions. The black dots (stars) are the $s$ ($d$)-wave interacting vertex.}
\label{Fig2}
\end{centering}
\end{figure}

\section{Atom-dimer scattering length}
The scattering length between an atom and a $d$-wave dimer can be obtained by calculating the scattering amplitude between them defined as $\mathcal{A}_{dd}({\bf k}, \omega, {\bf k}', \omega', E)$. In the subscript, the first letter $d$ stands for the incoming dimer to be $d$-wave type. The second letter $d$ denotes that the outgoing dimer is also a $d$-wave dimer. $\mathcal{A}_{dd}({\bf k}, \omega, {\bf k}', \omega', E)$ characterizes collision from an atom and a $d$-wave dimer to an atom and a $d$-wave dimer. For later convenience, definition of $\mathcal{A}_{sd}({\bf k}, \omega, {\bf k}', \omega', E)$ describes collision from an atom and an $s$-wave dimer to an atom and $d$-wave dimer. The incoming dimer is $s$-wave type and the outgoing dimer is $d$-wave type. Similar definition applies for $\mathcal{A}_{ss}({\bf k}, \omega, {\bf k}', \omega', E)$ or $\mathcal{A}_{ds}({\bf k}, \omega, {\bf k}', \omega', E)$.

The diagrams to calculate $\mathcal{A}_{dd}$ are shown in Fig. \ref{Fig2} \cite{Braaten}. In the center-of-mass frame, we take the external momentum of the atom and $d$-wave dimer to be $-{\bf k}$ and ${\bf k}$ for the incoming lines and $-{\bf k}'$ and ${\bf k}'$ for the outgoing lines. Here we only consider an on-shell case with $\omega^{(\prime)}=k^{(\prime)2}/2M$. In Fig. \ref{Fig2}, the double dashed lines are the bare $d$-wave dimer Green's function. The double solid lines are the bare $s$-wave dimer Green's function. The shaded double lines are the corresponding $s$- or $d$-wave dimer full propagator given by Eq. \eqref{EqDs} or \eqref{EqDd}. The black dots (stars) are the $s$ ($d$)-wave interacting vertex.

 On the right hand side of Fig. \ref{Fig2} (a), the lowest order of $\mathcal{A}_{dd}$ is the one-atom exchange term where the $d$-wave dimer is broken up to form another $d$-wave dimer. The second diagram includes all higher order of intermediate scattering process with both incoming and outgoing being the atom and $d$-wave-dimer, which gives the corresponding iterative Dyson equations. In the second diagram of Fig. \ref{Fig2} (a), within the one-atom exchange term, due to the existence of $s$-wave interaction, an $s$-wave dimer can be formed after the breaking of the incoming $d$-wave dimer. Then we need to include $\mathcal{A}_{sd}$ whose corresponding diagrams are shown in Fig. \ref{Fig2} (b). Similarly, the left two diagrams includes all higher order of scattering processes with both $d$-wave and $s$-wave interactions.  Due to the coexist of both $s$- and $d$-wave interactions, the $\mathcal{A}_{dd}$ and $\mathcal{A}_{sd}$ are coupled to each other as diagrammed in Fig. \ref{Fig2}, which form a coupled iterative Dyson equations. 

Explicitly, Fig. \ref{Fig2} can be expressed by two coupled STM equations for $\mathcal{A}_{dd}$ and $\mathcal{A}_{sd}$ which can be written as 
\begin{equation}\begin{aligned}
\int\frac{d^3{\bf k}_1}{(2\pi)^3}\left[\mathcal{M}_{dd}({\bf k}, {\bf k}_1)\mathcal{A}_{dd}({\bf k}_1, {\bf k}')+\mathcal{M}_{ds}({\bf k}, {\bf k}_1)\mathcal{A}_{sd}({\bf k}_1, {\bf k}')\right]& \\
-\mathcal{A}_{dd}({\bf k}, {\bf k}')=\mathcal{U}_{dd}({\bf k}, {\bf k}')& \\
\int\frac{d^3{\bf k}_1}{(2\pi)^3}\left[\mathcal{M}_{sd}({\bf k}, {\bf k}_1)\mathcal{A}_{dd}({\bf k}_1, {\bf k}')+\mathcal{M}_{ss}({\bf k}, {\bf k}_1)\mathcal{A}_{sd}({\bf k}_1, {\bf k}')\right]& \\
-\mathcal{A}_{sd}({\bf k}, {\bf k}')=\mathcal{U}_{sd}({\bf k}, {\bf k}')&.
\label{EqSTM}
 \end{aligned}\end{equation}
 On the right hand side of the above equations, the lowest scattering process $\mathcal{U}_{\sigma \sigma'}$ are given by 
 \begin{equation}\begin{aligned}
\mathcal{U}_{\sigma \sigma'}({\bf k}, {\bf k}')
=&U_{\sigma}\left(\frac{{\bf k}+2{\bf k}'}{2}\right)U_{\sigma'}\left( \frac{2{\bf k}+{\bf k}'}{2}\right) \\
&\times
G_0^{\rm A}\left({{\bf k}+{\bf k}'},E-\frac{k^2}{2M}-\frac{k'^2}{2M}\right)
\label{EqU}
\end{aligned}\end{equation}
where $U_{\sigma=s,d}$ are the $s$- or $d$-wave vertex with 
\begin{align}
&U_s({\bf k})=2e^{-k^2/\bar{\Lambda}_s^2}\\
&U_d({\bf k})=2\sqrt{2\pi}k^2Y_{20}(\hat{k})e^{-k^2/\bar{\Lambda}_d^2}, \label{EqUd}
\end{align} 
and 
\begin{equation}
G_0^{\rm A}({\bf k},\omega)=\frac{1}{\omega-k^2/2M+i0^+}
\end{equation}
is the atomic bare Green's function.
On the left hand side of the two coupled STM equations, the four kernels $M_{\sigma\sigma'=s,d}$ are given respectively by 
\begin{equation}\begin{aligned}
\mathcal{M}_{\sigma\sigma'}({\bf k}, {\bf k}_1)=\mathcal{U}_{\sigma\sigma'}({\bf k},{\bf k}_1)D_{\sigma'}\left({\bf k_1},E-\frac{k_1^2}{2M}\right) ,
\label{EqM}
\end{aligned}\end{equation}
in which $D_{\sigma=s,d}({\bf k},\Omega)$ is the $s$- or $d$-wave full Green's function of Eq. \eqref{EqDs} or \eqref{EqDd}.

\begin{figure}[t]
\begin{centering}
\includegraphics[width=0.4\textwidth]{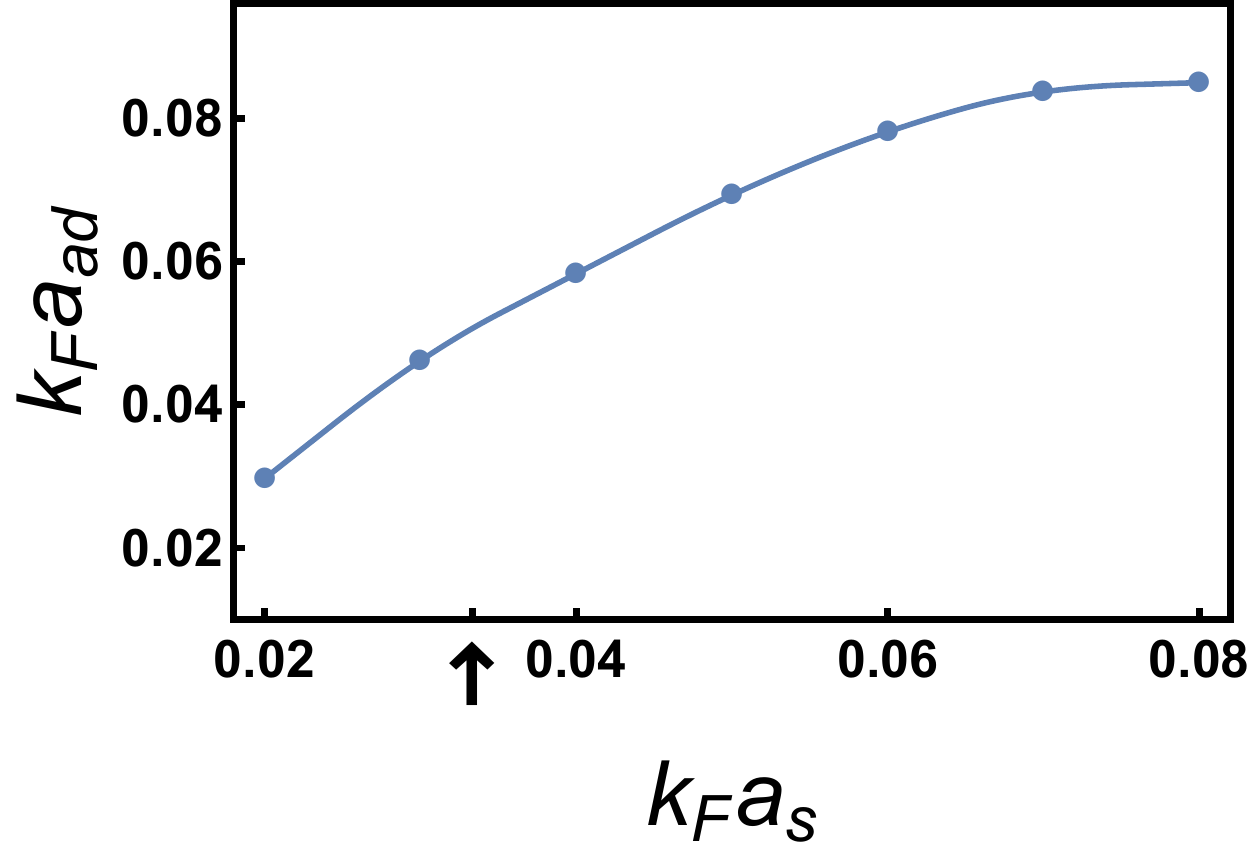}
\caption{The scattering length $a_{ad}$ between an atom and a $d$-wave dimer as a function of the atom-atom $s$-wave scattering length $a_s$. The $d$-wave bound state energy is fixed at $E_d=-0.5E_n$, where $E_n=k_n^2/2M$ with $k_n$ being the wave vector defined through the density $n=N/V$.  The arrow denotes the typical $a_s$ for the experimental observation of the $d$-wave Feshbach resonances \cite{XCYao} with $1/k_na_s=30$.}
\label{Fig3}
\end{centering}
\end{figure}

The information of low-energy collision between an atom and a $d$-wave dimer can be characterized by the atom-dimer scattering length which requires both the incoming and the outgoing momentum to be zero. Thus setting ${\bf k}'=0$ and projecting $\mathcal{A}_{d(s)d}({\bf k}, {\bf k}'=0)$ into components with total orbital angular momentum $l$, $\mathcal{A}_{d(s)d}({\bf k})\equiv\mathcal{A}_{d(s)d}({\bf k},{\bf k}'=0)$ can be written as
\begin{equation}\begin{aligned}
\mathcal{A}_{dd}({\bf k})=\sum_{l} \mathcal{A}_{dd}^l(k)P_l(\cos\theta_k)(2l+1),\\
\mathcal{A}_{sd}({\bf k})=\sum_{l} \mathcal{A}_{sd}^l(k)P_l(\cos\theta_k)(2l+1),
\end{aligned}\end{equation}
where $P_l(x)$ is the Legedre function with normalization condition $\int_{-1}^{1}dx P_l(x)P_{l'}(x)=2\delta_{ll'}/(2l+1)$. Then a partial wave expansion can be applied for the coupled STM equations Eq. \eqref{EqSTM}. In terms of $\mathcal{A}_{d(s)d}^l(k)$, Eq. \eqref{EqSTM} can be reduced into two coupled one-dimensional integral equations with integration over $k$ and summation over $l$, which we solve numerically.  In the numerical calculation, we discretize the integration of $k$ and reduce the two one dimensional integral equations into a matrix equation. A sufficient large cutoff for the momentum $k_c$ and angular momentum summation $l_c$ are chosen until the results of $\mathcal{A}_{d(s)d}^l(k)$ converge upon solving the reduced matrix equation. 

The atom and $d$-wave dimer scattering length is obtained through the $s$-wave component of the scattering $T$-matrix between an atom and a $d$-wave dimer with the energy $E$ fixed at the dimer binding energy $E_d$, i.e.  \cite{Braaten}
\begin{equation}\begin{aligned}
a_{ad}&=-\frac{M}{3\pi}T^{l=0}_{ad}({\bf k}={\bf k}'=0,E=E_d).
\end{aligned}\end{equation}
The atom and $d$-wave dimer $T$-matrix is related to the  calculated scattering amplitude through 
\begin{equation}
T_{ad}^{l=0}=Z_d \mathcal{A}^{l=0}_{dd},
\end{equation}
where $Z_d$ is the $d$-wave dimer residue factor and defined hrough the $d$-wave full Green's function when $E$ approaches the $d$-wave bound state energy as
\begin{equation}
D_d({\bf q}, E) \to \frac{Z_d}{E-E_\text{d}+i0^+},
\label{EqZd}
\end{equation}
where $E_\text{d}$ and $Z_\text{d}$ will be calculated numerically for each set of $\{ D, v, R\}$ according to Eq. \eqref{EqDd} and the dimer energy is approximated to $E_\text{d}\approx v/MD$ at the low energy limit.

The scattering length between an atom and a $d$-wave dimer $a_{\text{ad}}$ is plotted in Fig \ref{Fig3} as a function of the $s$-wave interaction strength between atoms for a fixed dimer energy.  It can be found the $a_{\text{ad}}$ is positive and increases with the increasing of atomic scattering length $a_\text{s}$, but its magnitude is much smaller than the result of mean-field approximation. Within the mean-filed theory, the $s$-wave interaction energy can be written in terms of total density of atoms, which reads  
\begin{equation}
\mathcal{E}=\frac{2\pi\hbar^2a_s}{M/2}\frac{(n_a+2n_d)^2}{2}
\end{equation}
where the cross term contributes to the atom-dimer interacting energy as
\begin{equation}
\mathcal{E}_{\text{ad}}=\frac{2\pi\hbar^24a_s}{M}n_an_d. \label{MF1}
\end{equation} 
Here $n_a$ is the atomic density and $n_d$ is the dimer density. 
Rewriting this energy in term of atom-dimer scattering length also by the mean-field approximation, it gives
\begin{equation}
\mathcal{E}_{ad}=\frac{2\pi\hbar^2a^{0}_{ad}}{2M/3}n_a n_d, \label{MF2}
\end{equation}
where $2M/3$ is the reduced mass of an atom and a dimer. 
By matching Eq. \ref{MF1} and \ref{MF2}, it yields $a_{\text{ad}}^0=8a_{s}/3$. As compared with the result shown in Fig. \ref{Fig3}, the actually three-body result is much smaller than the mean-field estimation.  {The deviation from the mean-field result also appears in other systems such as two-component Fermi gas \cite{aadSTM, PetrovFB} and Bose-Fermi mixture \cite{Ren}. For example, in the two-component Fermi gas, the atom-dimer scattering length $a_{ad}\sim1.23a_s$  \cite{Shin} which is much smaller than the mean filed result $8a_s/3$.}

\section{Three body Recombination rate}
\begin{figure}[t]
\begin{centering}
\includegraphics[width=0.35\textwidth]{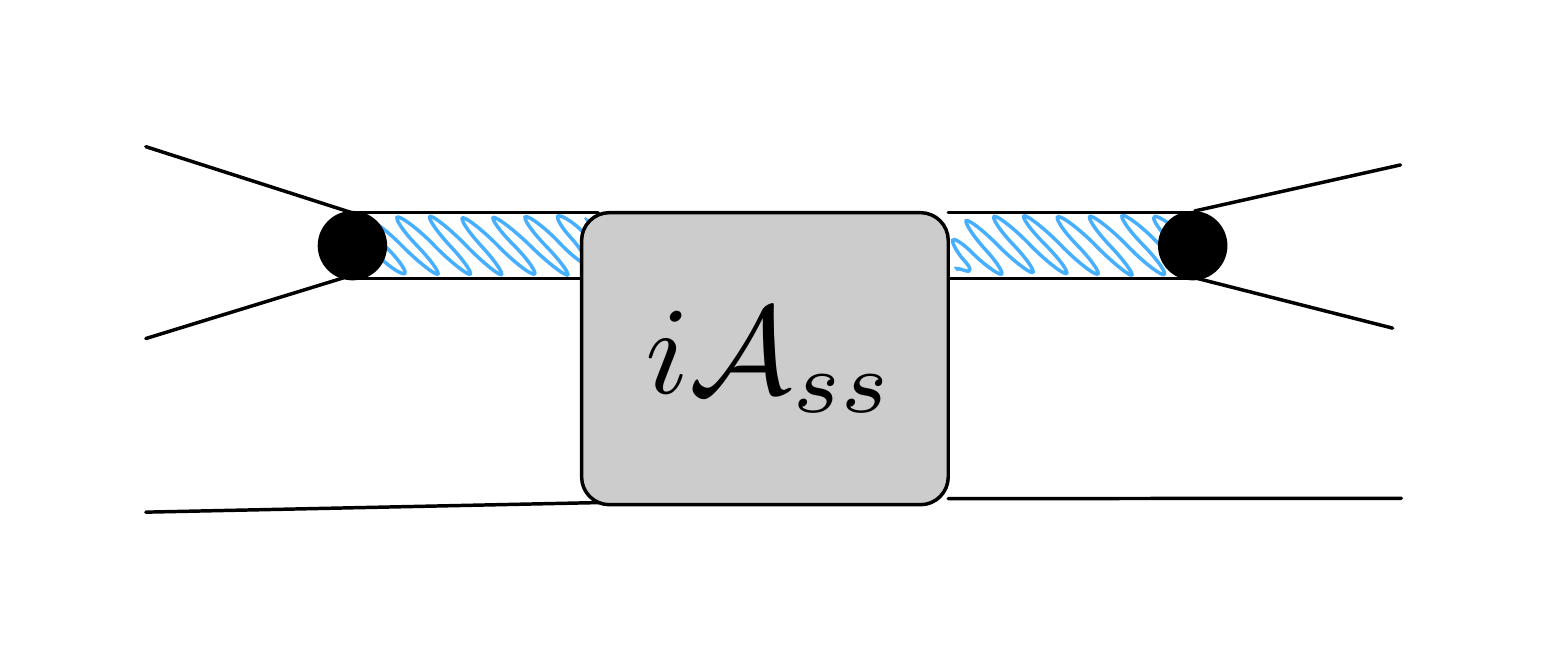}
\caption{Feynman diagrams of scattering matrix for three atoms $T_{aaa}({\bf k}=0, {\bf k}'=0,E=0)$.}
\label{Fig4}
\end{centering}
\end{figure}
\begin{figure}[t]
\begin{centering}
\includegraphics[width=0.5\textwidth]{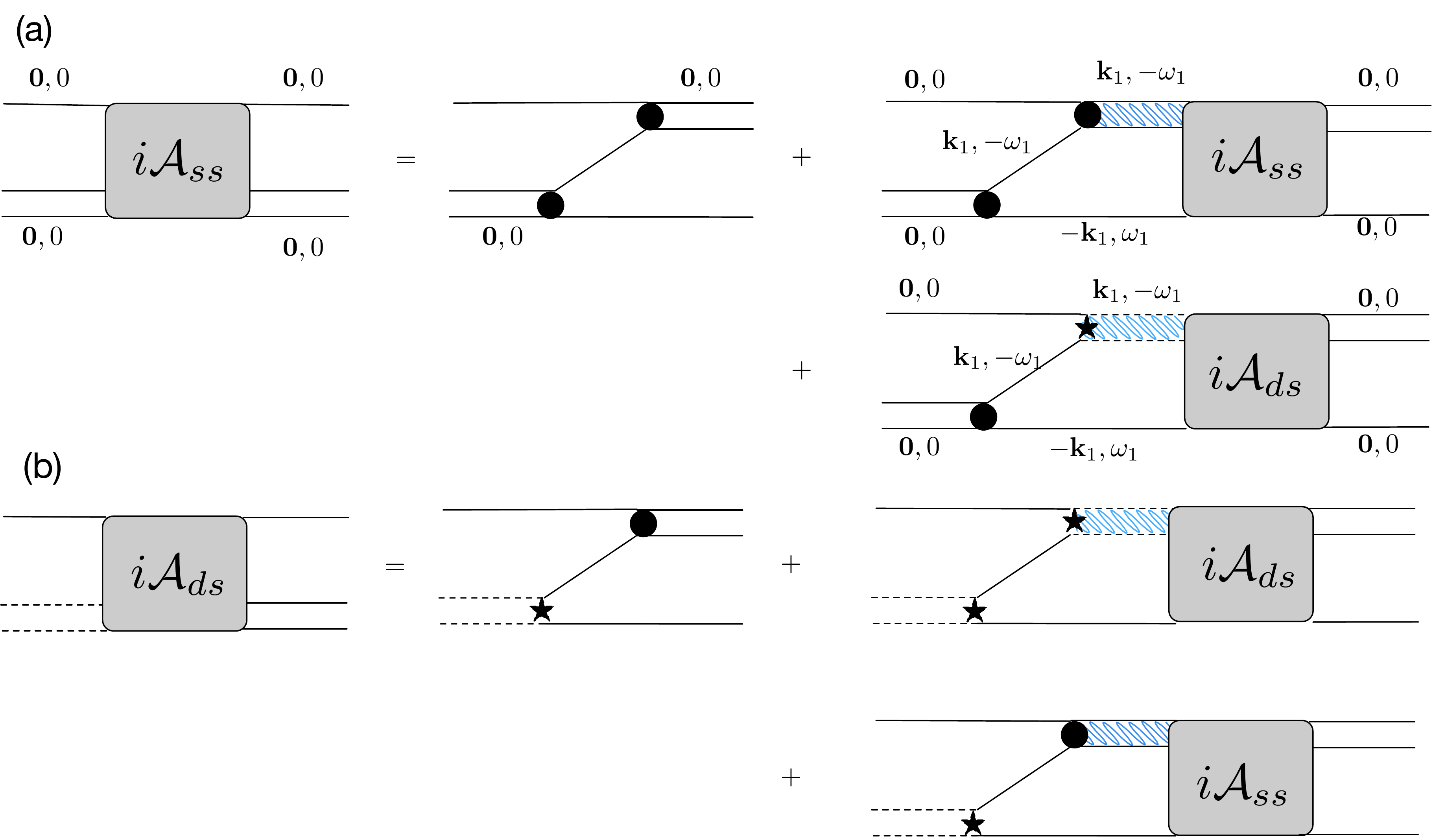}
\caption{Feynman rules for the scattering processes between an atom and an $s$-wave dimer. The double (dashed) lines are the bare $s$ ($d$)-wave dimer Green's functions. The shaded double lines are the corresponding full Green's functions. The black dots (stars) are the $s$ ($d$)-wave interacting vertex.}
\label{Fig5}
\end{centering}
\end{figure}

This formalism also allows us to study the three-body loss rate at low-energy, by considering collision of three atoms with nearly zero momentum. This calculation is very relevant to the lifetime of a low-temperature Bose condensate nearby such a $d$-wave resonance. As a result, different from considering the scattering between an atom and a $d$-wave dimer as discussed above, here we focus on the scattering process between three atoms with zero incoming momentum. 
In a condensate, almost all atoms condense at zero momentum and two zero-momentum atoms can only form $s$-wave dimer. Thus scattering of three zero-momentum atoms can be diagrammed as in Fig. \ref{Fig4} where only $s$-wave dimer is allowed to connect the zero-momentum atoms. Here, $\mathcal{A}_{ss}$ characterizes scattering between an atom and $s$-wave dimer with both the incoming and outgoing dimers being the $s$-wave type. As shown in Fig. \ref{Fig5} (a), the lowest order process of $\mathcal{A}_{ss}$ is also the one-atom exchange diagram. Although we consider a zero-momentum atom and $s$-wave dimer scattering process, in the intermediate scattering process finite momentum summation exists as indicated in the second and third terms. Thus similar to $\mathcal{A}_{dd}$, here $\mathcal{A}_{ss}$ is also coupled to $\mathcal{A}_{ds}$ with both $d$- and $s$-wave dimer formed during the intermediate scattering process. 

Without loss of generality, we consider a finite momentum scattering process and take the zero limit in the end.  Then the corresponding STM equations of Fig. \ref{Fig5} are written as 
\begin{equation}\begin{aligned}
\int\frac{d^3{\bf k}_1}{(2\pi)^3}\left[\mathcal{M}_{ss}({\bf k}, {\bf k}_1)\mathcal{A}_{ss}({\bf k}_1, {\bf k}')
+\mathcal{M}_{sd}({\bf k}, {\bf k}_1)\mathcal{A}_{ds}({\bf k}_1, {\bf k}')\right]&\\
-\mathcal{A}_{ss}({\bf k}, {\bf k}')=\mathcal{U}_{ss}({\bf k}, {\bf k}')& \\
\int\frac{d^3{\bf k}_1}{(2\pi)^3}\left[\mathcal{M}_{ds}({\bf k}, {\bf k}_1)\mathcal{A}_{ss}({\bf k}_1, {\bf k}')
+\mathcal{M}_{dd}({\bf k}, {\bf k}_1)\mathcal{A}_{ds}({\bf k}_1, {\bf k}')\right]&\\
-\mathcal{A}_{ds}({\bf k}, {\bf k}')=\mathcal{U}_{ds}({\bf k}, {\bf k}')& ,
\label{EqSTM2}
 \end{aligned}\end{equation}
where  $\mathcal{U}_{\sigma\sigma'}$ and the kernels $M_{\sigma\sigma'}$ are defined in Eq. \eqref{EqU} and \eqref{EqM}. Noticed that when taking the zero limit ${\bf k}={\bf k}'=E=0$, $\mathcal{A}_{ss}$ is infrared divergent due to the divergence of both $\mathcal{U}_{ss}$ and $\mathcal{M}_{ss}$ \cite{Petrov, BraatenSubv1, BraatenSubv2}. We seperate $\mathcal{A}_{ss}$ into the the regular part $\bar{\mathcal{A}}_{ss}$ and the irregular part as 
\begin{equation}\begin{aligned}
\mathcal{A}_{ss}({\bf k}, {\bf k}',E)=
&\bar{\mathcal{A}}_{ss}({\bf k}, {\bf k}',E)-\frac{4M}{k^2}+\frac{8M\pi a_s}{3k} \\
&-\left(\frac{8M\sqrt{3}a_s^2}{\pi}-\frac{256M\pi^3 a_s^2}{3}\right)\ln k,
\label{EqBarAss}
\end{aligned}\end{equation}
where the divergent irregular part comes from the lowest three order diagrams involving only $s$-wave dimer during the intermediate process  in Fig. \ref{Fig5} (a). Substituting Eq. \eqref{EqBarAss} into the coupled STM equations Eq. \eqref{EqSTM2}, the irregular part will cancel the divergence of $\mathcal{U}_{ss}$ and $\mathcal{M}_{ss}$.
In terms of regular part of the scattering amplitude $\bar{\mathcal{A}}_{ss}$, the STM equations Eq. \eqref{EqSTM2} is convergent and a similar numerical calculation can be adopted. 
\begin{figure}[t]
\begin{centering}
\includegraphics[width=0.45\textwidth]{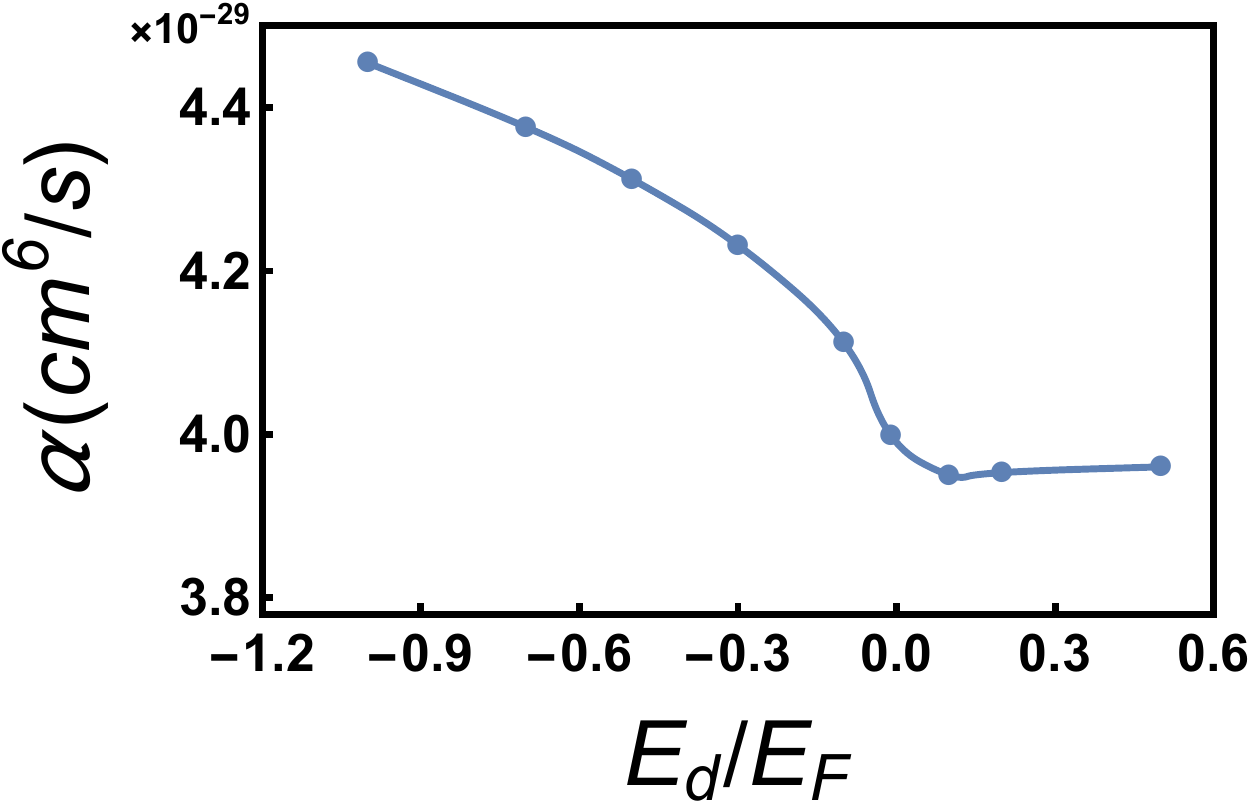}
\caption{Total three body recombination rate around the $d$-wave resonance at $1/k_na_s=30.$ In the weak coupling side with quasi-bound state, the loss is mainly due to the deep $s$-wave bound state. While in the strong coupling side, additional loss channel is the shallow $d$-wave bound state.}
\label{Fig6}
\end{centering}
\end{figure}

We solve the coupled STM equations Eq. \eqref{EqSTM2} with partial wave expansion in discrete momentum space. Finally, according to the optical theorem, the total three body recombination rate is related to the three-atoms scattering amplitude or the $\bar{\mathcal{A}}_{ss}$  through \cite{Braaten}
\begin{equation}
\alpha=\text{Im}[\mathcal{T}_{aaa}(0,0,0)]=\frac{144\pi^2{a}_s^2}{M^2}\text{Im}[\bar{\mathcal{A}}^{l=0}_{ss}(0,0,0)].
\label{EqalphaT}
\end{equation}
Noticed that, within the partial wave expansion, $l=0$ is the only term that contributes to three-body recombination rate $\alpha$ when ${\bf k}={\bf k}'=0$. In Fig. \ref{Fig6}, we plot the three-body recombination rate $\alpha$ as a function of $d$-wave dimer energy $E_d$. 
It can be found that $\alpha$ is small and almost uniform on the quasi-bound side. In this regime, atoms can escape from the trap mainly through colliding into a deep $s$-wave deep bound state and the tunneling between quasi-bound states and scattering states is negligible \cite{JPGaebler}. This is different from the finite temperature thermal gas case. For thermal atoms, they occupy an energy window with positive energy and peaked at $k_BT$. When the quasi-bound state energy is around $k_\text{B}T$, strong tunneling between quasi-bound  states and scattering states will result in a large loss in the quasi-bound side \cite{JPGaebler}. When the quasi-bound state energy is much larger than $k_\text{B}T$, the loss is suppressed. Thus, for thermal gas, the loss rate shows strong energy dependence at the quasi-bound side \cite{JPGaebler}. 

\begin{figure}[t]
\begin{centering}
\includegraphics[width=0.38\textwidth]{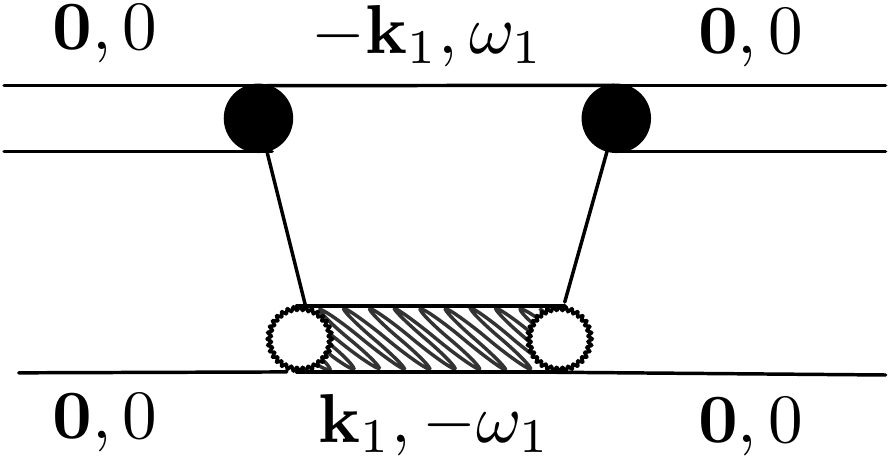}
\caption{The lowest Feynman diagram for scattering between the atom and $s$-wave dimer. The internal full dimer (grey shadowed double lines)could be $s$-, $p$-, or $d$-type. The connected two white circles are the associated vertex.}
\label{Fig7}
\end{centering}
\end{figure}

When approaching the bound side,  additional loss channel is open, in which  atoms can collide into shallow $d$-wave bound state. It can be found that the total recombination rate monotonically increases across the unitary point from the the quasi-bound to bound side as shown in Fig. \ref{Fig6}. To analytically understand the relation between the three-body recombination rate and the dimer bound state energy on the bound side, we consider the lowest order diagram for $\mathcal{A}_{ss}(0,0,0)$ as shown in Fig. \ref{Fig7}. For the $d$-wave interacting case, the full dimer propagator (grey shadowed double lines) in Fig. \ref{Fig7} stands for  $D_d({\bf k}_1,-\omega_1)$. We denote this lowest order $d$-wave type scattering amplitude as $\mathcal{A}_{ss}^{d}$, which is explicitly given by  
\begin{equation}
\begin{aligned}
\mathcal{A}_{ss}^{d}=\int\frac{ dk_1d\Omega }{(2\pi)^3} &k_1^2 U^2_s({\bf k}_1)  G_0^{\rm A}({\bf k}_1, -\frac{k_1^2}{2M})G_0^{\rm A}({\bf k}_1, -\frac{k_1^2}{2M}) \\
&\times U_d({\bf k}_1) U_d({\bf k}_1)   D_d\left({\bf k}_1,-\frac{k_1^2}{2M}\right).
\end{aligned}
\end{equation}
As given by Eq. \eqref{EqalphaT}, the three-body recombination rate is proportional to the imaginary part of  $\mathcal{A}_{ss}^d$ which is contributed by the pole of the $d$-wave full dimer propagator. Denoting the $d$-wave dimer bound state energy as $E_d=k_d^2/2M$, the imaginary part of the $d$-wave full dimer propagator Eq. \eqref{EqDd} can be written as
\begin{equation}
\begin{aligned}
\text{Im} \left[D_d\left({\bf k}_1,-\frac{k_1^2}{2M}\right)\right]
&=\frac{2M\pi Z_d}{3k_1}\delta(k_1-k_d),
\end{aligned}
\end{equation}
where $Z_\text{d}$ is defined by Eq. \eqref{EqZd}. We numerically find that $Z_d$ is nearly invariant through the resonance. Then the imaginary part of $\mathcal{A}_{ss}^d$ is evaluated as
 \begin{equation}
\begin{aligned}
\text{Im}[\mathcal{A}_{ss}^{d}]&=\int\frac{dk_1 d\Omega}{(2\pi)^3}  k_1^2U^2_s({\bf k}_1)  G_0^{\rm A}({\bf k}_1, -\frac{k_1^2}{2M})G_0^{\rm A}({\bf k}_1, -\frac{k_1^2}{2M}) \\
&~~~~~~~~~~~~~~\times U_d({\bf k}_1) U_d({\bf k}_1)   \frac{2M\pi Z_d}{3k_1}  \delta(k_1-k_d) \\
&\sim k_d^2 \cdot 1^2\cdot \frac{1}{k_d^2} \cdot \frac{1}{k_d^2} \cdot k_d^2 \cdot  k_d^2 \cdot \frac{1}{k_d}  \sim k_d,
\label{EqAAs}
\end{aligned}
\end{equation}
where we only count the order dependence on $k_d$. Then the recombination rate $\alpha_0^d$ due to the lowest $d$-wave scattering is proportional to $E_d$ with 
\begin{equation}
\alpha_0^d\sim \sqrt{E_d}.
\label{Eqalphad}
\end{equation}

When the intermediate dimer is $s$-wave type, the lowest scattering amplitude of $s$-wave type $\mathcal{A}_{ss}^s$ can be explicitly written as 
\begin{equation}
\begin{aligned}
\mathcal{A}_{ss}^{s}=\int\frac{dk_1d\Omega}{(2\pi)^3}  & k_1^2U^2_s({\bf k}_1) G_0^{\rm A}({\bf k}_1, -\frac{k_1^2}{2M})G_0^{\rm A}({\bf k}_1, -\frac{k_1^2}{2M}) \\
&\times U_s({\bf k}_1) U_s({\bf k}_1)   D_s\left({\bf k}_1,-\frac{k_1^2}{2M}\right),
\end{aligned}
\end{equation}
where 
\begin{equation}
\begin{aligned}
\text{Im} \left[D_s\left({\bf k}_1,-\frac{k_1^2}{2M}\right)\right]
&\sim\delta(k_1-k_s),
\end{aligned}
\end{equation}
with the $s$-wave dimer bound state energy $E_s=k_s^2/2M$. Then the three-body recombination rate due to lowest $s$-wave scattering is related to the corresponding $s$-wave bound state energy through 
\begin{equation}
\alpha_0^s\sim\text{Im}[\mathcal{A}_{ss}^{s}]\sim \frac{1}{k_s^2}\sim\frac{1}{E_s},
\end{equation}
where the enhanced coupling $k^2$ in Eq. \eqref{EqUd} due to $d$-wave interaction is missing. Here $\alpha_0^s\sim1/E_s$ is consistent to $\alpha_s\sim a_s^4$ upon including the prefactor $a_s^2$ given in Eq. \eqref{EqalphaT} \cite{Fedichev}. 
According to the above analysis, it turns out the $d$-wave interaction, due to the $k^2$ dependence of the interaction strength, the recombination rate $\alpha_0^d$ now is proportional  to $E_d$ which is consistent to the numerical calculation as shown in Fig. \ref{Fig6} on the bound side. This is very different from that for the $s$-wave interaction where the recombination rate is proportional to $1/E_s$. Thus for the $s$-wave case,  the loss peak is around the unitary point. 
For the $d$-wave interaction, the loss rate monotonically increase through the unitary point. Eventually this monotonic behavior will be suppressed by the finite momentum cut-off and  becomes less universal.

\section{Conclusion}
We study the three-body scattering processes in terms of the Skorniakov-Ter-Martirosion (STM) equations for the $d$-wave interacting Bose system with an $s$-wave background scattering. A theoretical formalism of the STM equations for mixed partial wave interactions is established. Within the STM equations, the resulting scattering length between an atom and a $d$-wave dimer provides the basic parameter for future investigation of this system. We calculate the total three-body recombination rate for the low-temperature Bose condensate. We find that the three-body recombination rate is almost a constant at the quasi-bound side, and this is strongly in contrast to a thermal gas nearby a high-partial wave resonance, where the recombination rate has strong energy dependence. We also find that the three-body recombination rate monotonically increases through the unitary point, and this is also strongly in contrast to a low-temperature gas nearby an $s$-wave resonance, where the recombination rate is peaked at unitary. This reveals that a combination of low-temperature degeneracy and $d$-wave resonance does yield new unique features.  

{\em Acknowledgement}. This work is supported MOST under Grant No. 2016YFA0301600  (HZ) and NSFC Grant No. 11734010 (HZ), Grants No. 11774426 (R.Q.), and the Fundamental Research Funds for the
Central Universities and the Research Funds of Renmin University of China under Grants No. 15XNLF18 (R. Q.) and No. 16XNLQ03 (R. Q.).

\end{document}